\documentclass[journal]{IEEEtran}
\usepackage[T1]{fontenc}
\usepackage[utf8]{inputenc}
\usepackage{amsmath,amssymb,booktabs,array,multirow,url,graphicx,cite,enumitem}
\usepackage{tikz}
\usetikzlibrary{positioning,arrows.meta,calc,fit}
\usepackage{xcolor}
\usepackage[hidelinks]{hyperref}
\newcommand{\tnote}[1]{\par\smallskip\raggedright\footnotesize #1\par}

\begin{document}

\title{Semantic Interoperability in Federated Digital Twin Systems: A Systematic Review
Through the Levels of Conceptual Interoperability Model}

\author{Manonmani Sekar and Nasim Nezamoddini%
\thanks{The authors are with the Department of Industrial and Systems
Engineering, Oakland University, Rochester, MI 48309 USA
(e-mail: manonmanisekar@oakland.edu). This work received no external funding. The
authors declare no competing interests.}}

\markboth{}%
{Sekar and Nezamoddini: Semantic Interoperability in Federated Digital Twin Systems}

\maketitle

\begin{abstract}
When two companies connect their digital twins, software models of their plants kept up to date with live data, each twin must translate the other's
data into its own terms, and that translation rule, called a mapping, can stop
being correct when one company changes the meaning of its data without telling the
other. This PRISMA 2020 systematic review screened 5,833 records and read 119 full
texts to determine whether any published system detects such a broken mapping and
repairs it while the twins operate. In all 32 included studies the mapping is
fixed before operation, and the two ingredients of run-time repair appear only
separately:
cross-company matching without any check that the match still holds, and detection and
repair of changed meaning within a single company. The review turns this gap into six
requirements, each illustrated on one running example, that a system claiming run-time
semantic interoperability between independently owned twins should meet.
\end{abstract}


\section{Introduction}\label{sec:intro}
\IEEEPARstart{A}{} digital twin is a virtual representation of a physical system, such as a machine, production line, or supply network, that is continuously updated using data from its physical counterpart. It enables the system to be monitored, its behaviour to be predicted, and operational decisions to be informed \cite{tao,kritzinger}. A manufacturer’s digital twin needs to coordinate with twins managed by its suppliers and energy provider. Each organisation maintains its own twin, so exchanging information and aligning operations becomes a shared challenge. This review  explores cross-organisational digital twins.

A supplier and an assembler each run a twin. The supplier's twin models the
supplier's plant and publishes how much it can produce, in a field called
\emph{capacity}, measured in units per eight-hour shift. The assembler's twin models
the assembly line and uses the incoming supply to plan production; it has no field
called capacity and plans instead with a \emph{throughput rate} in units per hour.
Before the connection goes live, an engineer writes a rule that divides the
supplier's capacity by eight and stores the result as throughput rate. That rule is a
\emph{mapping}: a correspondence stating which term on one side means
which term on the other, and how values convert between them. Months later the
supplier moves to twelve-hour shifts and, for its own reasons, redefines
\emph{capacity} as units per hour. Every message still parses, every value is still
plausible, and the assembler's plan is now wrong by a factor of eight.
It is an everyday example, yet it highlights an important underlying issue. Within a single company the
problem is settled by instruction, because one authority owns both models and can
change them together. Between companies there is no such authority. Each party
maintains its own model, discloses only part of it, and revises it on its own
schedule, so a mapping that was correct at deployment can become incorrect while
the system runs, and neither party is positioned to notice. Twins connected in this
way, each under a different owner, form what this review calls a \emph{federation}.

Research on connecting digital twins comes from three directions, and each one
handles a different step of the capacity example. The first is interoperability
standards, such as the Asset Administration Shell (AAS) and OPC UA, which fix how
data are structured and named \cite{wagner,ye}. Standards make sure that the
assembler's twin can receive the supplier's message and find the field called
\emph{capacity}. The second is ontologies and ontology matching, which describe
what each term means and find corresponding terms in two vocabularies
\cite{shvaiko,hogan}. They tell the assembler's twin that \emph{capacity} corresponds
to its own \emph{throughput rate}, which is the knowledge needed to write the
divide-by-8 mapping. The third is multi-agent reinforcement learning (MARL), in which
several decision-makers, for instance one per twin, learn by trial and error to
coordinate \cite{gronauer,canese}. It lets the assembler's planner adjust when
conditions on the shop floor change. Each direction therefore covers a step that
worked in the example: receiving the message, matching the terms, and planning the
line. None of them covers the step that failed, which is noticing that the
divide-by-8 mapping, correct when it was written, has stopped being correct.
When a supplier changes what its data means, its partners’ digital twins must recognise the change and adapt. This review uses the Levels of Conceptual Interoperability Model (LCIM) \cite{tolk2008,tolk2011} to assess that capability. Its six levels range from a working connection to shared model assumptions; level 5, dynamic interoperability, addresses changes in meaning during operation. Following PRISMA 2020 \cite{page2021}, this systematic review examines the published evidence to identify existing solutions or define the gap. Its central finding is that published approaches align the vocabularies of separately owned twins before operation, but none detects and repairs a mapping that becomes invalid while the twins are running.
The review asks three questions:
\begin{enumerate}[label=\textbf{RQ\arabic*},leftmargin=*]
    \item When is the mapping between digital twins established, and can it change during operation?
    \item What LCIM levels do standards, semantic technologies, and multi-agent learning support?
    \item What observable actions demonstrate that a system can repair a mapping at runtime?
\end{enumerate}

The paper makes four contributions:
\begin{itemize}[leftmargin=*]
    \item It defines mapping, drift, and repair through a running example and translates each LCIM level into a test for assessing studies.
    \item It identifies 5,833 records and finds that all 32 included studies establish mappings before operation; none repairs them at runtime.
    \item It shows that cross-company matching and runtime repair within one company have been demonstrated separately, but never together in one study.
    \item It defines six testable requirements for runtime repair between digital twins owned by different companies.
\end{itemize}

Section~\ref{sec:problem} defines the terms on the example and explains why learning
alone cannot solve the problem. Section~\ref{sec:method} describes how the literature
was searched and screened, Section~\ref{sec:results} reports what the included studies
show, and Section~\ref{sec:req} turns the gap into requirements.
Section~\ref{sec:discussion} compares this review with earlier ones and states its
limitations.

\section{The Problem}\label{sec:problem}

Digital twins owned by different companies rely on mappings to interpret exchanged information. When one company changes the meaning or units of its data without notifying the other, an existing mapping may continue to run but produce incorrect results. Such an unannounced change of meaning is called \emph{semantic drift}. Detecting it during operation, confirming its cause, and updating or suspending the affected mapping without redeploying the system is called \emph{repair}, and how to achieve repair is the problem this review examines.

\subsection{Federated Digital Twins}\label{sec:fed}

This review focuses on connected twins managed by independent parties, termed a \emph{federation} here. It uses four questions to identify relevant settings:
\begin{enumerate}[leftmargin=*]
    \item \emph{Independent control:} Can each party change its own twin without the other's approval?
    \item \emph{Operational independence:} Can each party continue its own operations if the connection is removed?
    \item \emph{Limited visibility:} Does each party expose only part of its model?
    \item \emph{Semantic dependence:} Must one twin interpret information defined by another party?
\end{enumerate}

The supplier and assembler meet all four criteria. Each controls its own twin and operates independently, but the assembler relies on capacity figures whose definition the supplier controls. Semantic dependence makes a mapping necessary. Independent control allows its meaning to change, while limited visibility makes that change harder to interpret. A mapping established before deployment may therefore become incorrect without stopping the exchange of data. The review excludes arrangements under a single controlling authority and studies that do not address interpretation across independently managed twins. It also distinguishes establishing a mapping before operation from maintaining its validity at runtime.

\subsection{LCIM Levels as Tests}\label{sec:lcim}

The review uses LCIM to assess interoperability
\cite{tolk2008,diallo2011}. Table~\ref{tab:lcim} translates each level into an observable test and illustrates it through the capacity example. These tests provide a consistent basis for assessing what each study demonstrates.
\begin{table*}[!t]\centering
\caption{LCIM levels as tests that can be applied to a paper, illustrated on the
supplier--assembler example.}\label{tab:lcim}\footnotesize
\begin{tabular}{@{}p{.07\textwidth}p{.30\textwidth}p{.55\textwidth}@{}}\toprule
Level & Test: the paper shows that\ldots & In the capacity example\\\midrule
L1 Technical & data travel between the twins over a working connection & The assembler receives the supplier's messages.\\
L2 Syntactic & the receiver parses the data with a shared format or information model & The assembler reads \emph{capacity} as a number in a known field of a known model.\\
L3 Semantic & the receiver interprets terms through a shared or formally mapped vocabulary & The assembler knows that \emph{capacity} corresponds to its own \emph{throughput rate}.\\
L4 Pragmatic & the receiver uses context such as units, time base, or provenance & The assembler knows the value is per eight-hour shift and converts it.\\
L5 Dynamic & a change of meaning during operation is detected, the mapping revised, and correct exchange restored & The supplier switches to units per hour; the assembler detects the change, replaces the conversion, and its plans become correct again.\\
L6 Conceptual & the parties reconcile their underlying model assumptions & The two firms agree what counts as available capacity, for instance whether maintenance time is included.\\
\bottomrule\end{tabular}
\tnote{Each study is coded at the highest level it demonstrates, not the level it
claims. Levels L1--L4 can all be met by an artefact that exists before the system
runs; L5 requires something to happen while it runs. A configurable mapping fixed at
design time is therefore L3 or L4, however flexible its interface, and a controller
that adapts its actions while the meaning of its inputs stays fixed does not reach
L5.}
\end{table*}

\subsection{Limits of Learning-Based Adaptation}\label{sec:formal}
A MARL controller, as introduced in Section~\ref{sec:intro}, learns a policy that
maximises a shared reward. It is tempting to expect such a controller to absorb drift
by adapting its behaviour. The capacity example shows why it cannot do so reliably.
After the supplier's change, the throughput rate that the assembler's planner receives
is eight times too small, yet it has the same form and range as before. Plans built on
it fall short, the reward drops, and the planner attributes the loss to its own
policy, which it adjusts to compensate for what is in fact a translation error. The
planner thus masks the drift rather than repairing it, and the divide-by-eight rule,
the object that actually changed, is never corrected.

The same argument can be stated formally. Let party $i$ express its observations in a
vocabulary $\Sigma_i$, and let party $j$ use the mapping
\begin{equation}
\phi_{ij}:\Sigma_i\rightarrow\Sigma_j\cup\{\bot\},\label{eq:mapping}
\end{equation}
where $\bot$ marks a term with no safe counterpart. Drift is the event
$\phi_{ij}^{(t)}\neq\phi_{ij}^{(t+1)}$ that occurs without notice to $j$; in the
example, $\phi_{ij}$ is the divide-by-eight rule, and drift is the supplier's switch
to units per hour. A cooperative learner under centralised training and decentralised
execution maximises
\begin{equation}
\mathbb{E}\!\left[\sum_{t=0}^{\infty}\gamma^{t}R(s_t,\mathbf{a}_t)\right],
\qquad a_{i,t}\sim\pi_i(\cdot\mid\tau_{i,t}),\label{eq:reward}
\end{equation}
over decentralised partially observable dynamics \cite{bernstein,oliehoek}. Nothing in
\eqref{eq:reward} refers to $\phi_{ij}$. The learner can change its policy $\pi_i$ but
not the mapping, so any loss that drift causes is charged to the policy.

\section{Review Methods}\label{sec:method}
\subsection{Protocol and Reporting}\label{sec:protocol}

The review aimed to identify systems demonstrating L5 interoperability between independently owned digital twins, using the test in Table~\ref{tab:lcim}, and to assess what other systems achieve.

The review is reported in accordance with the PRISMA 2020 statement \cite{page2021}, and the subsections below follow the corresponding PRISMA items: eligibility criteria, information sources and search strategy, study selection, and data collection and synthesis. The completed PRISMA 2020 checklist is provided with the supplementary
material (S5). The protocol, provided in Supplementary Material S1, was prepared on 16 September 2026, before the searches. It specified the eligibility criteria, search strings, full-text exclusion codes, LCIM tests, extraction variables, and agreement statistic. The review was not registered. Changes to the protocol, their reasons, and when they occurred are reported in Section~\ref{sec:limits}.

\subsection{Eligibility Criteria}\label{sec:eligibility}
At title and abstract, a study was eligible if it was a peer-reviewed journal article
or full conference paper in English, concerned digital twins, involved two or more
interacting twins, systems, or agents, used an interoperability, semantic, standards,
multi-agent, or learning mechanism, and gave enough technical detail to code. At full
text, eligibility was narrowed to the federation of Section~\ref{sec:fed}: two or more
twins or parties had to exchange data that the receiver must interpret. Full-text
exclusions carry one protocol code: E1 not a digital twin, E2 a single system or twin
without exchange between parties, E3 no relevant mechanism, E5 insufficient technical
detail, E6 wrong publication type, and E8 a preprint superseded by a peer-reviewed
version. The protocol defines ten codes, E1--E10 (S1); the four not listed here,
namely E4 nominal multi-agent only, E7 non-English, E9 full text unavailable, and
E10 outside the prespecified application or date scope, did not occur among the
full-text exclusions. For example, a factory that connects its own
machines to its own twin is excluded under E2, because no second party interprets its
data; the supplier--assembler connection would be included, because the assembler must
interpret terms that the supplier defines.

\subsection{Information Sources and Search Strategy}\label{sec:search}
Two sets of searches contribute records. In September 2026, Scopus, IEEE Xplore, and
EBSCO were searched for digital twins combined with multi-agent, learning, semantic,
and standards vocabulary; the Scopus query is reproducible and reported in
Supplementary Material S2, whereas
the IEEE Xplore and EBSCO queries were not recorded. On 24 October 2026, the three
protocol strings were run in OpenAlex over titles and abstracts \cite{priem2022},
restricted to English articles and reviews from 2015 to 2026. String A joins a
digital-twin block to interoperability, semantic, and standards terms, string B joins
it to multi-agent and decentralised-learning terms, and string C joins it to
federation and system-of-systems terms. The exact strings, filters, and run times are
in Supplementary Material S2. Records were deduplicated on DOI and then on
normalised title.

\subsection{Study Selection}\label{sec:selection}
Titles and abstracts were screened by a human reviewer and a machine reviewer against
the criteria of Section~\ref{sec:eligibility}. The first author screened all 5,833 unique records with Assistance of Large Language model. Agreement was computed on the 860 records and  the resulting statistic is reported in
Section~\ref{sec:agreement}. The second author then resolved the disagreements,
reading each record with both decisions and the model's stated reason in view; the
records she left undecided were resolved by re-reading them against the criteria with
the model's assistance. Because every machine decision carries its reason and is
released with the paper, each one can be checked by a reader. Open-access full texts
were retrieved through OpenAlex, Semantic Scholar, and arXiv; a preprint or repository
copy served only as access to the peer-reviewed report, which is the one cited. Full
texts were assessed against the eligibility criteria by a single reviewer with the
model's assistance, and each exclusion carries one of the protocol codes of
Section~\ref{sec:eligibility}. The screening flow is reported as a PRISMA 2020 flow
diagram in Fig.~\ref{fig:flow}, and every screening and full-text decision, with its
code and reason, is released with the supplementary material (S4).

\subsection{Data Collection, Appraisal, and Synthesis}\label{sec:data}
For each included study the reviewer recorded which parties exchange data, the
semantic artefact and alignment method, any agent or learner, the validation setting,
and, most importantly for RQ1, whether the mapping is fixed or can change during
operation (Supplementary Material S6). Following the protocol, studies were counted
as reports, and reports from the same project were linked so that their validation
evidence was not counted twice. Each study was assigned the highest LCIM
level it passes in
Table~\ref{tab:lcim}. Quality was appraised on eight protocol dimensions scored 0 to
2, covering problem clarity, realism of the federation, semantic and learning rigour,
agreement between metrics and claims, validation, reproducibility, and safety evidence
(Supplementary Material S7); appraisal describes the strength of evidence and excluded nothing. The synthesis
is narrative, because the studies share neither tasks nor metrics, and every count is
generated by script from the released decision files.

\section{Results}\label{sec:results}
\subsection{Selection and Screening Agreement}\label{sec:agreement}
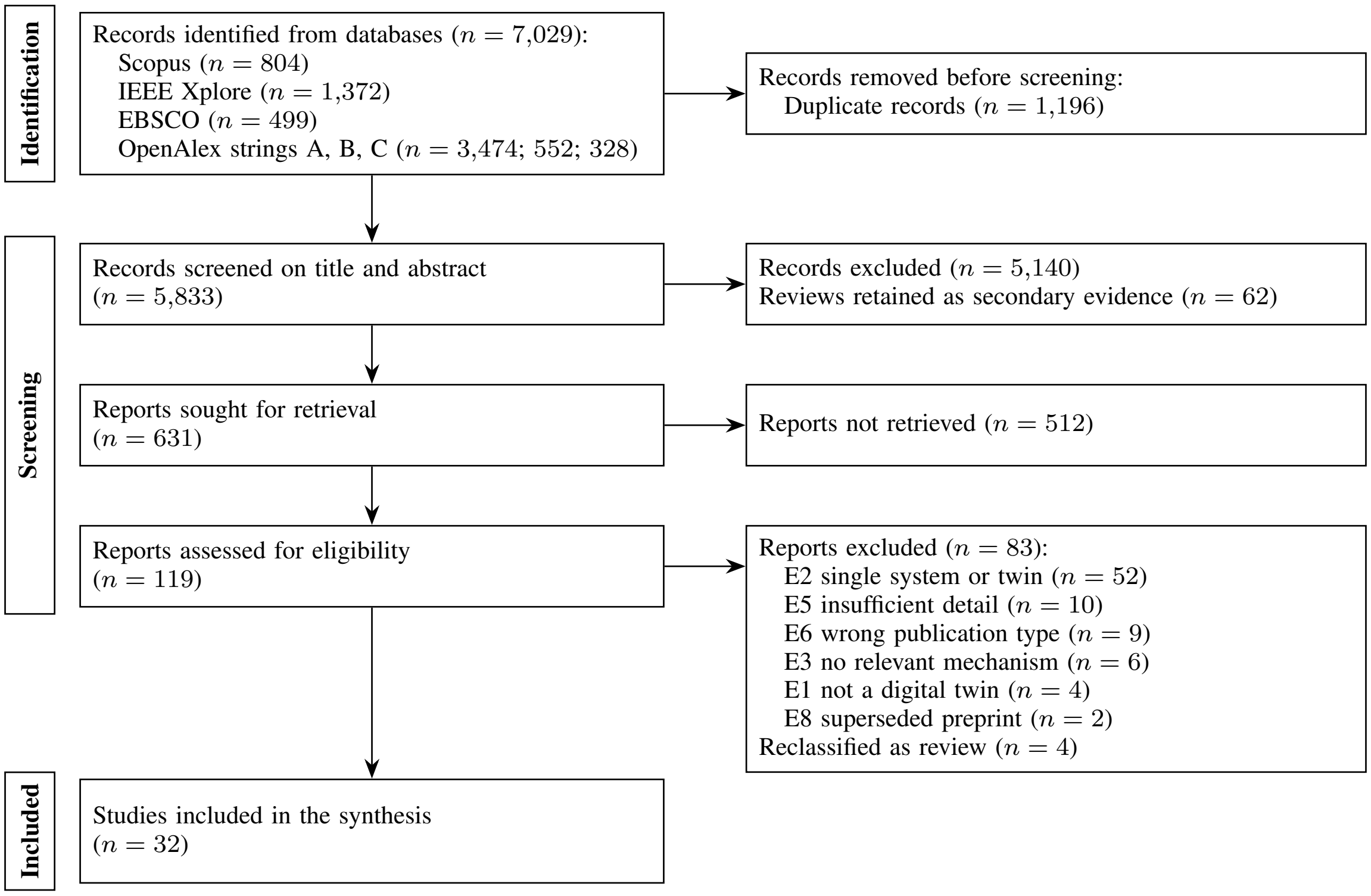
\begin{figure*}[!t]\centering\footnotesize
\begin{tikzpicture}[
  box/.style={draw=black, line width=0.5pt, rectangle, align=left, inner sep=4pt, text width=6.2cm, minimum height=0.9cm},
  sbox/.style={draw=black, line width=0.5pt, rectangle, align=left, inner sep=4pt, text width=6.6cm, minimum height=0.9cm},
  phase/.style={draw=black, line width=0.5pt, rectangle, minimum width=0.55cm, inner sep=2pt, font=\footnotesize\bfseries},
  arr/.style={-{Stealth[length=2.2mm]}, line width=0.5pt}]
\node[box] (id) at (0,0) {Records identified from databases ($n=7{,}029$):\\
\quad Scopus ($n=804$)\\
\quad IEEE Xplore ($n=1{,}372$)\\
\quad EBSCO ($n=499$)\\
\quad OpenAlex strings A, B, C ($n=3{,}474$; $552$; $328$)};
\node[sbox, anchor=west] (dup) at ($(id.east)+(0.9,0)$) {Records removed before screening:\\
\quad Duplicate records ($n=1{,}196$)};
\node[box, below=0.75cm of id] (scr) {Records screened on title and abstract\\($n=5{,}833$)};
\node[sbox, anchor=west] (exc) at ($(scr.east)+(0.9,0)$) {Records excluded ($n=5{,}140$)\\
Reviews retained as secondary evidence ($n=62$)};
\node[box, below=0.65cm of scr] (sou) {Reports sought for retrieval\\($n=631$)};
\node[sbox, anchor=west] (nr) at ($(sou.east)+(0.9,0)$) {Reports not retrieved ($n=512$)};
\node[box, below=0.65cm of sou] (ass) {Reports assessed for eligibility\\($n=119$)};
\node[sbox, anchor=north west] (ftx) at ($(ass.north east)+(0.9,0)$) {Reports excluded ($n=83$):\\
\quad E2 single system or twin ($n=52$)\\
\quad E5 insufficient detail ($n=10$)\\
\quad E6 wrong publication type ($n=9$)\\
\quad E3 no relevant mechanism ($n=6$)\\
\quad E1 not a digital twin ($n=4$)\\
\quad E8 superseded preprint ($n=2$)\\
Reclassified as review ($n=4$)};
\node[box, below=1.9cm of ass, minimum height=1.15cm] (inc) {Studies included in the synthesis\\($n=32$)};
\draw[arr] (id) -- (scr); \draw[arr] (scr) -- (sou); \draw[arr] (sou) -- (ass); \draw[arr] (ass) -- (inc);
\draw[arr] (id) -- (dup); \draw[arr] (scr) -- (exc); \draw[arr] (sou) -- (nr); \draw[arr] (ass.east) -- (ass.east -| ftx.west);
\path (id.west) ++(-0.55,0) coordinate (px);
\node[phase, anchor=center, fit={(px |- id.north) (px |- id.south)}] (p1) {};
\node[rotate=90, font=\footnotesize\bfseries] at (p1) {Identification};
\node[phase, anchor=center, fit={(px |- scr.north) (px |- ass.south)}] (p2) {};
\node[rotate=90, font=\footnotesize\bfseries] at (p2) {Screening};
\node[phase, anchor=center, fit={(px |- inc.north) (px |- inc.south)}] (p3) {};
\node[rotate=90, font=\footnotesize\bfseries] at (p3) {Included};
\end{tikzpicture}
\caption{PRISMA 2020 flow diagram of study selection \cite{page2021}.}\label{fig:flow}
\end{figure*}

Fig.~\ref{fig:flow} traces the selection. Agreement between the two screeners was low,
with Cohen's $\kappa=0.15$, a chance-corrected statistic on which 1 means perfect
agreement and 0 means agreement no better than chance (raw agreement 0.58), on the 860 records screened by both.
The cause is identifiable rather than random. Most records that the first pass kept
and the second reviewer dropped describe one twin with an ontology or a learning
method but no second party; the second reviewer applied the multi-party condition at
the title stage, whereas the first pass had deferred it to full text. Of the 362 disagreements, the second author resolved 248, siding with the first pass
in 147 and with the second reviewer in 101; the remaining 114 were resolved with
language-model assistance. At full
text the same reason dominated the exclusions: many studies presented as
interoperability work turn out to connect systems inside one organisation, where a
single authority can prescribe the terms.

\subsection{Characteristics of Included Studies}
Table~\ref{tab:included} lists the 32 included studies. They come from 31
projects, because Akroyd \emph{et al.} and Bai \emph{et al.} both report The World
Avatar and are treated as one source of validation evidence
\cite{akroyd2021,bai2024}. Two further pairs share authors but describe different
projects, and each of their reports is counted separately. They span manufacturing, cities,
energy, water, agriculture, biodiversity, health, shipping, nuclear decommissioning,
and supply chains, and almost all appeared from 2022 onward. Most connect parties under
separate authority, such as municipal departments, research infrastructures, or
supply-chain partners \cite{wolf2022,ranatunga2025}. The rest connect several twins
inside one organisation or on one shared platform \cite{elshafei2025,gil2024}. The
evidence is mostly a prototype, a worked example, or a simulation, and the two-site
self-driving laboratory of Bai \emph{et al.} is the only study that runs a closed loop
between organisations on real equipment \cite{bai2024}. Read against the capacity
example, a typical row such as Stamer \emph{et al.}, in which a manufacturer shares AAS
submodels with its network partners \cite{stamer2022}, would let the assembler parse
the supplier's capacity field (L2). The meaning of that field, and any later change to
it, would still rest on prior agreement.

\begin{table*}[!t]\centering
\caption{Included studies coded at the highest LCIM level demonstrated.}\label{tab:included}\scriptsize
\setlength{\tabcolsep}{3pt}
\begin{tabular}{@{}p{.15\textwidth}p{.25\textwidth}p{.27\textwidth}p{.19\textwidth}c@{}}\toprule
Study & Parties exchanging information & Semantic artefact or mechanism & Validation & LCIM\\\midrule
Liu \emph{et al.} \cite{liu2026watershed} & Heterogeneous watershed platforms & Ontology; embedding-based matching & Prototype on case data & L3\\
Akram \emph{et al.} \cite{akram2026} & Medical twins at separate institutions & Federated MARL; model updates only & Simulated ICU & L2\\
Ricci \emph{et al.} \cite{ricci2022} & Linked twins of many owners & RDF knowledge graph & Illustrative use cases & L3\\
Xu \emph{et al.} \cite{xu2023asc} & Supply-chain organisations & Agent negotiation protocol & Food supply-chain prototype & L2\\
Akroyd \emph{et al.} \cite{akroyd2021} & Cross-domain agents, one platform$^\dagger$ & Dynamic knowledge graph & Use cases & L3\\
Elshafei \emph{et al.} \cite{elshafei2025} & Robots of two vendors, one lab$^\dagger$ & AAS submodels & Lab experiments & L2\\
Raes \emph{et al.} \cite{raes2022} & City models of several providers & Common data model via broker & Pilot cities & L2\\
Kallenberg \emph{et al.} \cite{kallenberg2025} & Crop and disease twins, RL agents$^\dagger$ & FIWARE NGSI-LD data models & Two field pilots & L2\\
Santhanavanich \emph{et al.} \cite{santhanavanich2025} & OGC pilot participants & OGC API standards & Interoperability pilot & L2\\
Gonz\'alez-Gerpe \emph{et al.} \cite{gonzalez2024} & Lower-level twins aggregated$^\dagger$ & WoTDT ontology; mapping aggregation & Worked example & L3\\
Gil \emph{et al.} \cite{gil2024} & Coupled component twins$^\dagger$ & Semantic lifting; consistency queries & Two case studies & L3\\
Jacques \emph{et al.} \cite{jacques2025} & Tools of European partners & Common decommissioning ontology & Facility demonstration & L3\\
Mathews \emph{et al.} \cite{mathews2023} & Stations and AGVs, one system$^\dagger$ & Ontology-generated models; VDA 5050 & Demonstrator line & L2\\
Bai \emph{et al.} \cite{bai2024} & Laboratories in two countries & Ontologies; dynamic knowledge graph & Real hardware, two sites & L3\\
Moretti \emph{et al.} \cite{moretti2023} & Built-environment data owners & Federated open data model & Campus case & L3\\
Vakil \emph{et al.} \cite{vakil2026} & LCA twins feeding city twins & Unit and boundary harmonisation & Conceptual only & L4$^\ast$\\
Wolf \emph{et al.} \cite{wolf2022} & Emergency services and city & Shared cloud data model & Prototype use case & L2\\
Ranatunga \emph{et al.} \cite{ranatunga2025} & Water authorities' databases & Ontology-based data access & Lake use case & L3\\
Islam \emph{et al.} \cite{islam2026} & Research infrastructures & RO-Crate metadata, identifiers & Ten prototypes & L2\\
Herzog \emph{et al.} \cite{herzog2027} & City departments and stakeholders & OGC API standards & Deployed platform & L2\\
Fonseca \emph{et al.} \cite{fonseca2023} & Ship-lifecycle stakeholders & Mapping to multiple taxonomies & Two case studies & L3\\
Stamer \emph{et al.} \cite{stamer2022} & Manufacturer and network partners & AAS submodels & Worked example & L2\\
Padsala \emph{et al.} \cite{padsala2026} & Municipal departments & CityGML with Energy ADE & District case study & L3\\
Gorissen \emph{et al.} \cite{gorissen2026} & Institutes sharing robot data & Annotated digital shadows & Proof of concept & L3\\
Ou \cite{ou2025} & Transnational supply-chain nodes & Carbon-factor database & Simulation & L2\\
Paul and Goodridge \cite{cr1526} & Federated multi-domain twins & Ontology alignment; evolution proposed & Simulation & L3\\
Moshrefzadeh \emph{et al.} \cite{moshrefzadeh2020} & Landscape data owners & Metadata model and catalogue & Research-centre demonstration & L2\\
Vergara-Marcillo \emph{et al.} \cite{vergara2026} & Pre-existing twins of separate stakeholders & Schema mediation in a federation node & Prototype, emergency response & L2\\
Tsampras \emph{et al.} \cite{tsampras2025} & Asset and context providers (urban) & NGSI-LD descriptor service & Proof of concept & L2\\
Kannappan \emph{et al.} \cite{kannappan2025} & Lifecycle organisations (product passport) & AAS data on a blockchain & Latency and throughput tests & L2\\
Hua \emph{et al.} \cite{hua2026} & Electric vehicles of separate owners & MADDPG with twin-assisted critic & Simulation & L2\\
Alvi \emph{et al.} \cite{alvi2026} & City traffic twin and external data providers & FIWARE NGSI-LD data model & Deployed city case & L2\\
\bottomrule\end{tabular}
\tnote{$^\dagger$ Several twins within one organisation or on one common platform;
all other rows involve parties under separate authority. $^\ast$ L4 by design; no
exchange is demonstrated. Full extraction and appraisal are in S6 and S7.}
\end{table*}

\subsection{Timing of Mapping Establishment}\label{sec:fixed}
The answer to RQ1 is uniform. In every included study the mapping between parties is
established before the twins operate, whether written by hand, agreed in a project
ontology, defined by a common standard, or computed once by a matcher. Applied to
the capacity example, every one of these systems would carry the divide-by-eight rule into
operation and keep applying it after the supplier's change. A few studies allow
something to change at run time, but in each case the change stops short of the
mapping. The World Avatar knowledge graphs add and update instances while their
ontologies stay fixed \cite{akroyd2021,bai2024}. Gil \emph{et al.} check at run time
that coupled twins remain consistent, yet leave any repair to an engineer
\cite{gil2024}. Paul and Goodridge propose evolving the ontologies of federated twins
but do not run the mechanism \cite{cr1526}.

The LCIM levels follow from this. No study passes the L5 test of
Table~\ref{tab:lcim}. Fewer than half reach L3, in which the receiver interprets terms
through a shared ontology or knowledge graph, and almost all the others stop at L2,
sharing a format through AAS submodels, OGC APIs, FIWARE data models, schema mediation,
or a broker's common model. L4, the explicit use of units and context, appears only in
a design: Vakil \emph{et al.} specify a translation layer that harmonises functional
units between life-cycle-assessment twins and city twins, without demonstrating it
\cite{vakil2026}. Separate ownership changes the level reached little: 15 of the 26
studies across organisations stop at L2, against 3 of the 6 inside one organisation.

Appraisal explains why these results cannot be read more generously. The studies
describe their problem and their federation well, but few report a metric that would
show whether data were interpreted correctly, few release code or data
\cite{kallenberg2025,bai2024}, and safety evidence is almost absent. A reader of most
included papers therefore cannot tell whether interpretation succeeded even before
drift occurs.

\subsection{Contributions of Standards, Semantics, and Learning}\label{sec:families}
Answering RQ2 by body of work shows a clear division of labour. Standards carry the
L2 studies: they guarantee that the assembler can parse the supplier's message, and
they reach L3 only where the semantic identifiers and dictionaries they reference are
jointly governed \cite{idta,iec62541}. Semantic technologies carry the L3 studies:
they make the correspondence between \emph{capacity} and \emph{throughput rate}
explicit, and their alignments are judged, when they are judged at all, by their
quality at the moment of deployment \cite{shvaiko,olala}. Multi-agent learning is
rare. The two federated multi-agent learners exchange only privacy-protected model
updates or high-level shared state, with no vocabulary to align
\cite{akram2026,hua2026}. In Kallenberg \emph{et al.} reinforcement learning operates
over a fixed FIWARE interface \cite{kallenberg2025}. No included study combines a
learner with an explicit cross-party mapping, so the masking effect described in
Section~\ref{sec:formal} has not been examined empirically.

\subsection{Partial Approaches to Repair}\label{sec:halves}
The nearest approaches to L5 come from two directions, and they do not meet. On the
cross-company side, matching between separately owned twins is computed, but only
once. Liu \emph{et al.} map heterogeneous watershed platforms to a unified ontology by
embedding similarity when the platforms are integrated \cite{liu2026watershed}. Paul
and Goodridge align the ontologies of federated twins and propose a mechanism for
their evolution, which they do not run \cite{cr1526}. In the capacity example either
approach would find that \emph{capacity} corresponds to \emph{throughput rate}, and
neither would revisit the match after the supplier's change. The nearest run-time
approach found by the search, by Amadeo \emph{et al.}, was excluded at title-and-abstract
screening. It matches service names between twins when a lookup fails, but it caches
matches without ever invalidating them \cite{amadeo2026}, so it too would keep applying
the divide-by-eight rule.

On the repair side, two studies detect a change of meaning and correct an ontology
while the system runs, but each does so inside a single company. Abbasi \emph{et al.}
detect semantic drift between the layers of one model-driven twin and realign its
metamodel and ontology \cite{abbasi2026mgmt}. Globa and Ushakov update ontology modules
when a system's ontology drifts away from its own data stream, with an expert
approving each update \cite{globa2026}. Earlier agent architectures built on semantic
models likewise fix their mappings within one party before execution
\cite{cr048,cs01924}. The capability that the capacity example needs, detecting and
repairing a broken mapping between two companies, thus exists in two halves held by
different studies. A search of the remaining eligible abstracts for vocabulary of
change, versioning, and repair found none describing cross-company repair.

\section{Requirements for Run-Time Repair}\label{sec:req}
\subsection{Six Requirements}
RQ3 asks what a system would have to do, observably, to repair the capacity mapping.
Table~\ref{tab:req} answers with six requirements, each derived from a gap above and
each stated so that it can be tested regardless of the technique used to meet it.

\begin{table*}[!t]\centering
\caption{Requirements for run-time repair of cross-party mappings, each shown on the
capacity example.}\label{tab:req}\footnotesize
\begin{tabular}{@{}p{.035\textwidth}p{.15\textwidth}p{.33\textwidth}p{.39\textwidth}@{}}\toprule
ID & Requirement & What it demands & In the capacity example\\\midrule
R1 & Explicit mapping & The mapping $\phi_{ij}$ is a run-time object that records its target, confidence, provenance, validation status, and the model versions on both sides. & The rule ``\emph{capacity} $\div 8$'' is stored with the supplier model version it was written for, so a version change can be seen.\\
R2 & Representable non-translation & A term that cannot be translated safely is marked $\bot$ and is distinguishable from a real value. & While the rule is in doubt, the assembler receives ``capacity unknown'' rather than a wrong number.\\
R3 & Drift detection & A change in meaning is detected from observed data, without notice from the other party, within the planning horizon. & The assembler notices that reported capacity has jumped relative to delivered quantities.\\
R4 & Repair under partial disclosure & A new mapping is found without either party exposing its whole model, and governance states who may approve it. & The assembler asks the supplier only for the time base of \emph{capacity}, and the supplier's answer is logged before the new rule goes live.\\
R5 & Separated credit assignment & A learner can tell a translation error from a policy error, so that \eqref{eq:reward} stays well posed under drift. & The scheduler does not retrain its policy to compensate for an eightfold unit error.\\
R6 & Interoperability-specific evaluation & Benchmarks inject unannounced changes of meaning and report how quickly and correctly they are repaired. & The test suite switches \emph{capacity} to units per hour mid-run and measures time to detect, time to repair, and plans made on wrong data.\\
\bottomrule\end{tabular}\end{table*}

The requirements differ in difficulty. R1 and R2 ask only that information a system
already holds be kept rather than discarded, and their absence reflects habit more
than difficulty. R3, R4, and R5 are open research problems, and R6 is shared
infrastructure that the field must build collectively; the companion paper addresses
it for one architecture \cite{partB}.

\subsection{Architectural Implications}
The requirements point towards a division of responsibility. Mappings, their
provenance, their validation status, and the $\bot$ marker are naturally symbolic
objects, whereas spotting an anomaly in delivered quantities and proposing a candidate
conversion suit learned components \cite{garcez}. A neuro-symbolic arrangement would
let a learned component propose a new mapping $\widehat{\phi}_{ij}$ while a symbolic
component decides whether to adopt it:
\begin{equation}
\operatorname{Accept}(\widehat{\phi}_{ij})\iff
\operatorname{Valid}_{\mathrm{sem}}\wedge
\operatorname{Safe}_{\mathrm{op}}\wedge
\operatorname{Evidence}(\widehat{\phi}_{ij})\geq\eta .
\end{equation}
In the example, a proposed ``\emph{capacity} $\div 1$'' rule would be adopted only if
it is semantically valid, safe for the plans that depend on it, and supported by
enough evidence. Safety adds a further reason for this separation, since a safety
shield presumes that the predicates it checks mean what their designer intended
\cite{alshiekh}, which is exactly what drift undermines. The included studies show the
symbolic half to be mature and its coupling to learning to be untested, so this
arrangement is a design implication rather than an established result.

A knowledge base is a natural home for the symbolic half, provided that each mapping
is stored as a record rather than embedded in code. The Simple Standard for Sharing
Ontological Mappings already describes a mapping's source, target, justification, and
confidence \cite{sssom}. PROV-O records who proposed a mapping and on what evidence
\cite{provo}, and SHACL expresses the validation rules that a proposed mapping must
satisfy before it is adopted \cite{shacl}, so the acceptance rule above can be
evaluated as constraint checks over the stored record. In the capacity example the
knowledge base would hold the record ``\emph{capacity} $\div 8$'', stamped with the
supplier model version it was written for, its provenance, and the status
\emph{validated}; once the drift detector raises doubt, the status becomes
\emph{suspect} and the assembler receives $\bot$ instead of a number. Such a store
meets R1 and R2 directly and supports the governance part of R4. It cannot meet R3 by
itself, because a knowledge base records what is known and does not notice that the
supplier's meaning has changed. Ownership matters as well. A single knowledge base
holding both companies' semantics would restore the central authority that a
federation excludes, so each party should keep its own knowledge base and share only
the mapping records, for instance through a neutral registry or a data-space
connector.

\subsection{Evaluation of Repair Claims}
A convincing demonstration of L5 would report four things that no included study
reports. It would inject controlled, unannounced changes of meaning of several kinds,
such as changed units, changed aggregation, split or merged concepts, and new concepts,
each logged with the correct mapping before and after. It would measure
interoperability directly, through the rate of invalid translations, the time to
detect and to repair, and the share of usable exchanges before drift, after drift,
and after repair, alongside the usual task metrics. It would let parties exchange only
targeted evidence, such as units or example values, rather than complete models. And
it would keep a trace of every mapping's versions and decisions, because two systems
can earn the same reward while making very different semantic decisions. A control
horizon $H$ should be declared, with
$\Delta t_{\mathrm{detect}}+\Delta t_{\mathrm{repair}}\leq H$, so that a slow offline
procedure is not reported as run-time repair.

\section{Discussion}\label{sec:discussion}
\subsection{Comparison with Earlier Reviews}
Earlier reviews cover digital-twin concepts, data integration, knowledge graphs in
twins, and semantic interoperability in Industry 4.0 (Table~\ref{tab:prior}). Lehmann
\emph{et al.} come closest and independently observe that implementations lack
closed-loop control \cite{cr511}. None of them distinguishes independently owned twins
from twins under one authority, applies a single interoperability scale across
standards, semantics, and learning, or asks what happens when a meaning changes after
deployment. Those three choices are what allow the present review to locate the gap
precisely.

\begin{table*}[!t]\centering
\caption{This review compared with earlier ones. ``Owners'' separates independently
owned from single-authority twins; ``LCIM'' organises findings by interoperability
level; ``Change'' examines what happens when a meaning changes after deployment.}
\label{tab:prior}\small
\begin{tabular}{@{}p{.23\textwidth}p{.29\textwidth}p{.17\textwidth}ccc@{}}\toprule
Review & Scope & Method & Owners & LCIM & Change\\\midrule
Liu \emph{et al.} \cite{liudt} & DT concepts and applications & Narrative review & No & No & No\\
Piroumian \cite{piroumian} & DT interoperability & Position article & Partly & No & No\\
Hildebrandt \emph{et al.} \cite{hildebrandt2024} & Data integration for DTs & Syst.\ review & No & No & No\\
Listl \emph{et al.} \cite{listl2024} & Knowledge graphs in DTs & Syst.\ review & No & No & No\\
Huebner and Ihlenfeldt \cite{huebner2026} & Semantic interop.\ in I4.0 & Mapping study & No & No & No\\
Gebhard and Wang \cite{gebhard2026} & Semantic interop.\ maturity & Maturity model & No & Adjacent & No\\
Bacnak \emph{et al.} \cite{bacnak2026} & Cognitive digital twins & Syst.\ review & No & No & No\\
Lehmann \emph{et al.} \cite{cr511} & Semantic enrichment with MAS & Syst.\ review & No & No & No\\
Gronauer and Diepold \cite{gronauer} & Multi-agent deep RL & Syst.\ survey & n/a & No & No\\
\midrule
\textbf{This review} & \textbf{Alignment between independently owned DTs} &
\textbf{PRISMA syst.\ review} & \textbf{Yes} & \textbf{Yes} & \textbf{Yes}\\
\bottomrule\end{tabular}
\end{table*}

\subsection{Barriers to Repair}
Repair is hard partly for technical reasons. Standards, ontologies, and learners work
on different objects, namely message formats, vocabularies, and policies, and a
changing mapping falls between them. Behaviour alone cannot reveal whether a partner's
data changed because its plant changed, its policy changed, or its meaning changed,
which is why a system expected to repair meaning, rather than compensate for it, needs
an explicit set of hypotheses about what a term might now mean.

It is also hard for organisational reasons. Where the included studies connect
separate parties, the mapping was agreed in a project, a pilot, or a standard, and that
agreement is what makes the connection work at all. Run-time repair asks the assembler
to reinterpret the supplier's data without such an agreement, which turns a semantic
question into a governance one: who proposed the new mapping, on what evidence, under
which model versions, and who may approve or revoke it. The consequences are
asymmetric, too. An assembler that holds ``capacity unknown'' can pause or ask,
whereas one that holds a wrong number plans with confidence.

\subsection{Limitations}\label{sec:limits}
Only open-access full texts were assessed, and the findings refer to the 119 reports
read in full. The review departs from its protocol in five further ways. OpenAlex
replaced the protocol's subscription databases for the frozen strings, Web of Science
and the ACM Digital Library were not searched, EBSCO was searched although the
protocol had not listed it, and two of the September queries were
not recorded. The research questions were also reframed: the protocol's five
questions were consolidated into the three reported here, with the definition of a
federated twin (protocol RQ1) addressed in Section~\ref{sec:fed} and the questions
on integration patterns, evidence for dynamic interoperability, and remaining gaps
(protocol RQ3--RQ5) folded into RQ1--RQ3. Finally, the second reviewer at title and abstract was a language model. It alone
screened 4,973 records, and 114 of the 362 disagreements were resolved with its
assistance rather than by the second author. Full texts were assessed by a single
reviewer with the same assistance. These choices made a full screen of 5,833 records
feasible. Agreement on the dual-screened records was low; its cause is traceable to
the stage at which one criterion was applied, and the coefficient is reported as
computed. Every decision is released so that each step can be inspected and repeated.

No requirement is new in isolation, since versioning, ontology evolution, provenance,
and constrained communication are established topics. The contribution lies in
showing, on systematic evidence, that they have not been combined for independently
owned twins, and in stating what combining them would have to demonstrate. A
neuro-symbolic design is one way to meet the requirements; a probabilistic symbolic
system or a constrained language-model matcher could meet them equally.

\section{Conclusion and Further Work}
When two companies connect their digital twins, the published systems agree on how
to translate each other's data before the connection goes live, and they keep that
translation unchanged afterwards. If the supplier quietly redefines \emph{capacity},
none of the 32 included systems would notice. The ingredients of a remedy exist, in
cross-company matching on one side and single-company drift repair on the other, but
no study joins them. The six requirements and the evaluation design set out here state
what joining them would have to show.

Three directions follow. A shared benchmark with controlled, unannounced changes of
meaning and published reference mappings would let competing designs be compared on
repair rather than on task reward alone. Repair under partial disclosure, in which
parties exchange targeted evidence instead of complete models, deserves study as a
problem in its own right. Finally, extending the full-text assessment to
subscription-only work and to the databases not covered here would show whether the
gap reported in this review persists; the sooner a study closes it, the better.

\section*{Supplementary Material}
Released with this paper: the protocol (S1); the search strings, filters, dates,
and exports (S2); the deduplicated record set with source attribution (S3); every
title-and-abstract and full-text decision with its code and reason, the disagreement
resolutions, and the retrieval logs (S4); the PRISMA 2020 checklist (S5); the
extraction sheet (S6); the quality appraisal (S7); and the analysis scripts and
README that generate every count reported here (S8). All of these materials are
publicly available as supplementary files accompanying this article.

\section*{Acknowledgment}
A large language model (Anthropic Claude) assisted 
assessment, extraction, wrote the retrieval and counting scripts, and
assisted language editing. Every decision it recorded is released with the
supplementary material, and the authors take full responsibility for the content.

\begin{IEEEbiographynophoto}{Manonmani Sekar}
received the B.E. degree in electronics and communication engineering from Madurai
Kamaraj University, India, in 1992, and the M.E. degree in embedded systems from
Anna University, Chennai, India, in 2014. She is currently working toward the Ph.D.
degree in systems engineering with the Department of Industrial and Systems
Engineering, Oakland University, Rochester, MI, USA, where she is a research
assistant in the Complex Systems Laboratory. Before her doctoral studies she spent
more than two decades in engineering practice in manufacturing and
telecommunications. Her research interests include semantic interoperability in
federated digital twin systems, multi-agent reinforcement learning, and safe
coordination of multi-vendor industrial automation.
\end{IEEEbiographynophoto}

\begin{IEEEbiographynophoto}{Nasim Nezamoddini}
is with the Department of Industrial and Systems Engineering, Oakland University,
Rochester, MI, USA. Her research interests include multi-agent systems, intelligent
manufacturing, and digital twins.
\end{IEEEbiographynophoto}

\end{document}